\documentclass[11pt]{article}
\usepackage{cite,epsf,amsmath,amssymb}
\newcommand{\dd}{{\rm d}}
\newcommand{\ddoverdd}[1]{{\dd\over \dd #1}}
\newcommand{\doverd}[1]{{\partial\over \partial #1}}
\newcommand{\order}[1]{{\cal O}(#1)}

\newcommand{\msbar}{\overline{\mbox{\abbrev {MS}}}}

\newcommand{\remark}[1]{{\center
    \fbox{\parbox{14cm}{\footnotesize\sf #1}}\\[1em]}}
\renewcommand{\remark}[1]{}
\newcommand{\eqn}[1]{Eq.\,(\ref{#1})}
\newcommand{\norm}{{\cal N}}
\newcommand{\code}{\footnotesize\sf}
\newcommand{\GFermi}{G_{\mbox{{\tiny F}}}}
\newcommand{\abbrev}{\small\sc}
\newcommand{\ep}{\epsilon}
\newcommand{\ponetwo}{p_1\cdot p_2}
\newcommand{\apib}{{\alpha_s^\bare\over \pi}}
\newcommand{\api}{{\alpha_s\over \pi}}
\newcommand{\bare}{{\mbox{\tiny B}}}
\newcommand{\lag}{{\cal L}}
\newcommand{\cala}{{\cal A}}

\newcommand{\lmt}{l_t}
\newcommand{\mtop}{M_\mathrm{t}}
\newcommand{\MHiggs}{M_\mathrm{H}}
\begin{document}
\title{
  \begin{flushright}
    \sf\normalsize BNL-HET-00/29,
    \sf\normalsize hep-ph/0007289 ---
    \sf\normalsize July 2000\\[3em]
  \end{flushright}
  Virtual corrections to $gg\to H$ to two loops\\ in the heavy top limit
  }
\author{
  Robert V.~Harlander\\[1em]
  \it\normalsize HET, Physics Department\\
  \it\normalsize Brookhaven National Laboratory, Upton, NY 11973
  }
\date{}
\maketitle 
\begin{abstract}
  The virtual corrections to the production cross section of a Standard
  Model Higgs boson are computed up to order $\alpha_s^4$. Using an
  effective Lagrangian for the limit $\mtop \to \infty$, we evaluate the
  relevant massless two-loop vertex diagrams by mapping them onto
  three-loop two-point functions, following a method recently introduced
  by Baikov and Smirnov~\cite{BSvert}.  As a result, we find a
  gauge-invariant contribution to the total Higgs production cross
  section at {\footnotesize NNLO}.
\end{abstract}
                                

\section{Introduction}

The Standard Model of Elementary Particle Physics ({\abbrev SM}) has
been verified in many of its details with enormous precision over the
last 20~years.  However, several fundamental questions remain unanswered.
Perhaps the most important unresolved problem concerns the origin of the
particle masses.

In the {\abbrev SM}, the underlying mechanism for mass generation is
spontaneous breakdown of the electro-weak gauge symmetry. In its minimal
version (which we will assume throughout this paper) it predicts a
single, as yet undetected physical particle, the Higgs boson.  It is
determined to be electrically neutral and of spin zero. Its mass,
however, is a free parameter of the theory.

The extensive searches for the Higgs boson at particle colliders have
set a lower bound on its mass, the latest update yielding a limit of
around 108\,GeV~\cite{lepewwg}.  On the other hand, theoretical
predictions for physical observables depend on the Higgs mass through
radiative corrections.  Comparison of the existing experimental
precision data with their theoretically predicted values allows to derive
a most probable range for the Higgs mass which turns out to be roughly
between 100 and 200\,GeV~\cite{lepewwg}.

With {\abbrev LEP} being close to its maximum possible energy, the
attention concerning Higgs search is turning towards future experiments
at hadron colliders, in particular {\abbrev LHC}, scheduled
for the year 2007, or already {\abbrev Tevatron}'s Run\,{\abbrev II},
starting in 2001.

The dominant production mechanism for a Higgs boson with a mass below
1\,TeV at the {\abbrev LHC} will be through gluon-gluon fusion (for a
review see~\cite{spira}). The coupling of the gluons to the Higgs boson
is mediated through a quark loop, Fig.\,\ref{dia::triangle}\,(a).  In
the heavy quark limit, the corresponding form factor becomes independent
of the quark mass.  Thus, this process can be used, for example, to
count the number of heavy quarks that may exist beyond the third
generation.

The current theoretical prediction for this reaction carries an
uncertainty of about a factor of 1.5 to 2.  It is therefore important to
improve on the theoretical accuracy.  In this paper we provide a gauge
invariant ingredient to the complete next-to-next-to-leading order
prediction, namely the virtual corrections up to order $\alpha_s^4$.
The calculation is, to our knowledge, the first application of a
recently introduced method that allows to relate the relevant set of
vertex diagrams to the more familiar class of three-loop two-point
functions.


\section{Effective Lagrangian}

\begin{figure}
  \begin{center}
    \leavevmode
    \begin{tabular}{cc}
      \epsfxsize=4.cm
      \epsffile[120 240 510 480]{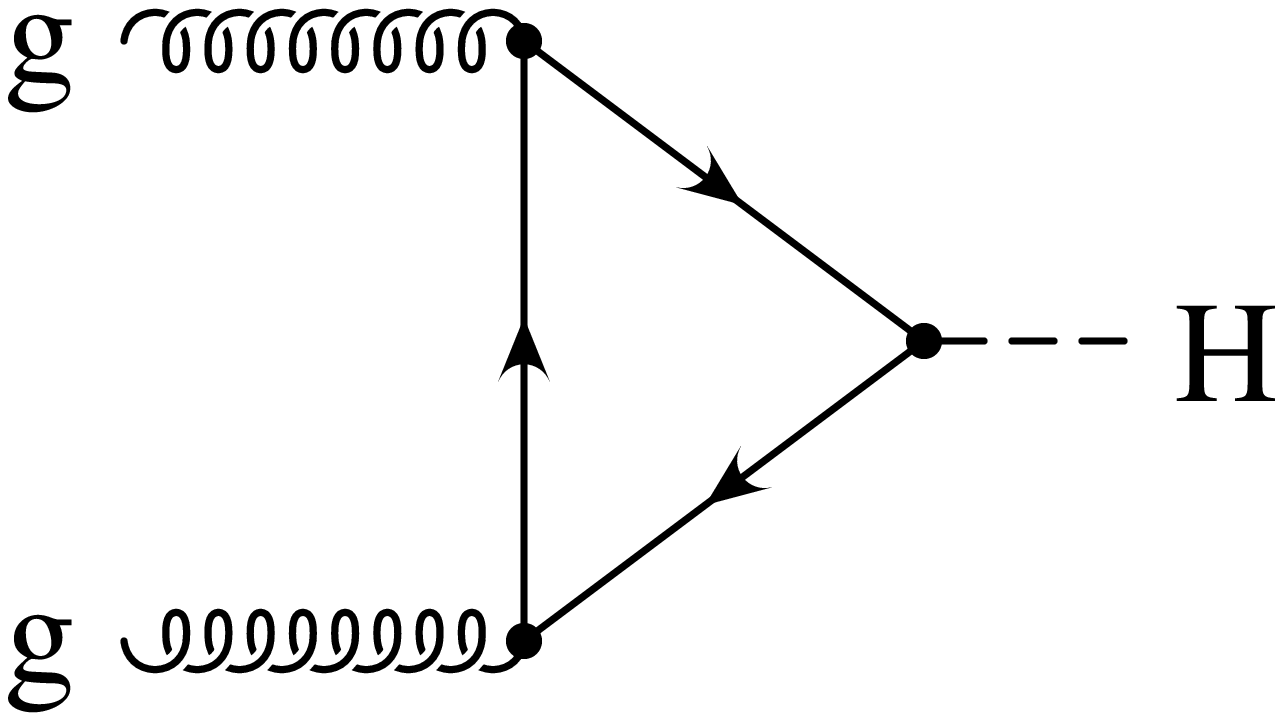} &\qquad
      \epsfxsize=4.cm
      \epsffile[120 240 510 480]{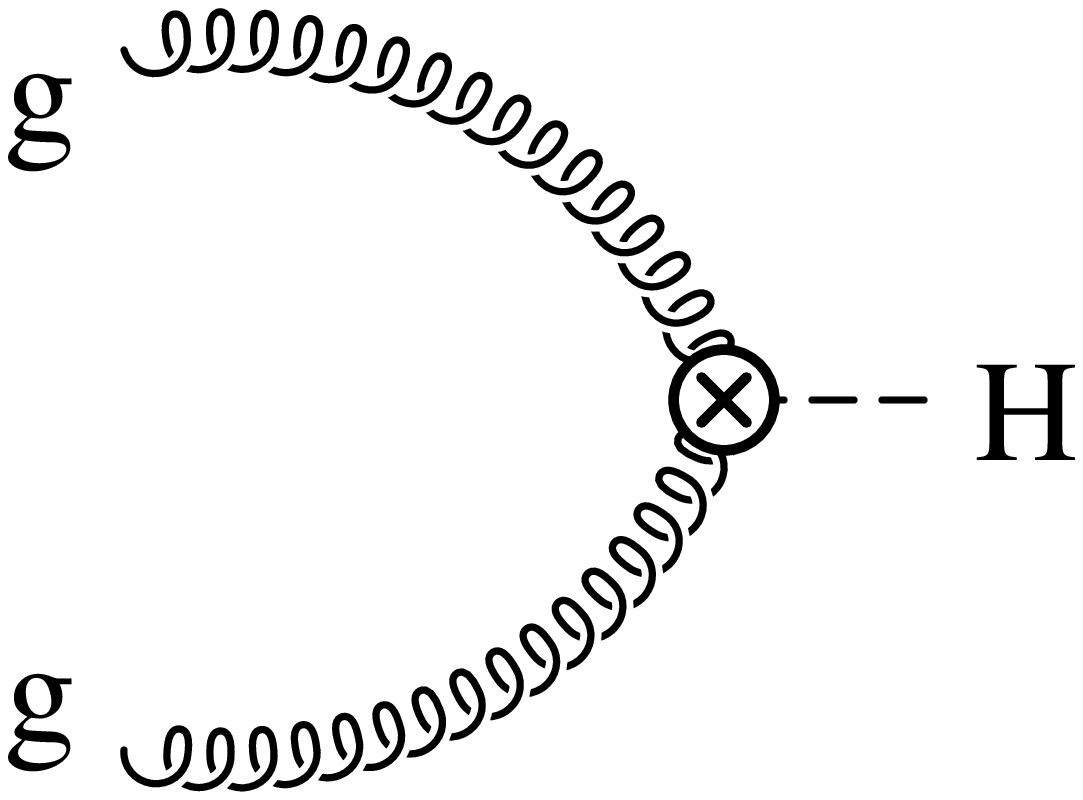}\\
      (a) & (b)
    \end{tabular}
    \parbox{14.cm}{
      \caption[]{\label{dia::triangle}\sloppy
        Leading order diagram to the process $gg\to H$: (a)~in the full and
        (b)~in the effective theory. The
        ``$\otimes$'' denotes the effective vertex of Eq.\,(\ref{eq::leff}).
        }}
  \end{center}
\end{figure}

As it was mentioned before, the coupling of the gluons to the Higgs
boson is mediated through a quark loop, Fig.\,\ref{dia::triangle}\,(a).
Since all quarks except for the top are much lighter than the current
lower limit on the Higgs mass, we will neglect their masses in the
following.  In this case, the top quark is the only one that couples
directly to the Higgs boson, because the Higgs-fermion vertex is
proportional to the fermion mass.
The leading order result has been known for quite a while \cite{lo}. At
the parton level it reads:
\begin{equation}
  \begin{split}
    \sigma_{\mbox{\tiny LO}}(gg\to H) &= {\GFermi\alpha_s^2(\mu^2)\over
      128 \sqrt{2}\pi}\,\tau^2\,\delta(1-z)\,
    \left|1+(1-\tau)f(\tau)\right|^2\,,\\[1em]
    f(\tau) &= \left\{
      \begin{array}{ll}
        \arcsin^2{1\over \sqrt{\tau}}\,, & \tau\ge 1\,,\\
        -{1\over 4}\left[\log{1+\sqrt{1-\tau}\over 1-\sqrt{1-\tau}} -
          i\pi\right]^2\,,&\tau < 1\,,
      \end{array}
    \right.
    \\[1em]
    \tau &= {4\mtop^2/\MHiggs^2}\,,\qquad 
    z = {\MHiggs^2/s}\,,
    \label{eq::lo}
  \end{split}
\end{equation}
where $s$ is the partonic cms energy and $\GFermi$ is the Fermi coupling
constant. $\alpha_s$ is the strong coupling constant which depends on
the renormalization scale $\mu$. $\mtop$ is the pole mass of the top
quark, and $\MHiggs$ is the Higgs mass. In order to arrive at the cross
section for hadron collisions, $\sigma_{\mbox{\tiny LO}}$ has to be
folded with the gluon distribution functions.

The currently favored values for the Higgs mass appear to be not too
different from the top quark mass. However, since the threshold for open
top production is at $\MHiggs = 2\mtop$, an expansion in terms of small
Higgs mass is expected to work well for $\MHiggs < 2\mtop$.  In fact,
for the most interesting mass range of 100\,GeV\,$\lesssim \MHiggs
\lesssim$\,200\,GeV it appears that at next-to-leading order in
$\alpha_s$ the complete result is excellently approximated
by the leading term in an expansion in $\MHiggs^2/\mtop^2$
\cite{dawson,kauffman1}.  Therefore we think it is reasonable to adopt
this limit also at next-to-next-to-leading order ({\abbrev NNLO}).

All calculations in this paper have been performed using dimensional
regularization in $D=4-2\ep$ space time dimensions.  Unless stated
otherwise, renormalized expressions are to be understood in the $\msbar$
scheme, bare ones will be marked by the superscript~``{\small B}''.
Furthermore, all the quantities used in the following refer to the five
flavor effective theory. For example, by $\alpha_s$ we mean the running
coupling constant with five active flavors, $\alpha_s^{(5)}(\mu^2)$.

The most convenient way to obtain the leading term in
$\MHiggs^2/\mtop^2$ is to use the following effective Lagrangian for the
Higgs-gluon interaction, where the top quark has been integrated out:
\begin{equation}
\lag_\mathrm{eff} = -{H\over v} C_1^\bare\,{1\over 4} 
(G_{\mu\nu}^\bare)^2
= -{H\over v} C_1\,{1\over 4} (G_{\mu\nu})^2\,.
\label{eq::leff}
\end{equation}
Here, $G_{\mu\nu}$ is the gluonic field strength tensor in the effective
five flavor theory, and $v$ is the vacuum expectation value of the Higgs
field, related to the Fermi constant by $v = (\sqrt{2}\GFermi)^{-1/2}$.
The coefficient function $C_1$ has been computed in \cite{CheKniSte97}
up to $\order{\alpha_s^4}$.  In order to obtain the cross section for
$gg\to H$ with {\abbrev NNLO} accuracy, however, it will only be needed
up to $\order{\alpha_s^3}$~\cite{c1a3}:
\begin{equation}
\begin{split}
C_1 =& -{1\over 3}\api\bigg\{
  1+{11\over 4}\api + \left(\api\right)^2\bigg[
  {2777\over 288} + {19\over 16}\lmt +
  n_l\bigg(-{67\over 96} + {1\over 3}\lmt\bigg)\bigg]\bigg\}\,,
\label{eq::coefc}
\end{split}
\end{equation}
\remark{Note that here we already use $\alpha_s^{(5)}$ as mentioned
  above, as opposed to \cite{CheKniSte97}. The formula is calculated in
  \mbox{\tt \~\//math/ggh/CheKniSte.m}.}
where $l_t=\ln(\mu^2/M_t^2)$, with $M_t$ the on-shell top quark mass.
Here and in the following, the number of (light) flavors $n_l$ will
eventually be set to five.

The renormalized operator in Eq.\,(\ref{eq::leff}) is related to the bare
one through~\cite{spiridonov}
\begin{equation}
(G_{\mu\nu})^2 = {1\over 1-\beta(\norm\alpha_s)/\ep}(G_{\mu\nu}^\bare)^2\,,
\end{equation}
where
\begin{equation}
\norm = \exp\left[\ep\left(-\gamma_{\mbox{\tiny E}} + \ln 4\pi\right)\right]\,,
\end{equation}
and $\beta(\alpha_s)$ governs the running of the strong coupling constant:
\begin{equation}
\mu^2\ddoverdd{\mu^2}\alpha_s = \alpha_s\beta(\alpha_s)\,.
\end{equation}
Its perturbative expansion is known up to
$\order{\alpha_s^4}$~\cite{beta4}, but for the present purpose the
terms up to $\order{\alpha_s^{2}}$ are sufficient:
\begin{equation}
\begin{split}
  \beta(\alpha_s) =& -\api\bigg[
  {1\over 4}\left(11 - {2\over 3}n_l\right)
  +\api{1\over 16}\left(102 - {38\over 3}n_l\right)\bigg]\,.
  \label{eq::beta}
\end{split}
\end{equation}
Using the Lagrangian of Eq.\,(\ref{eq::leff}) for the Higgs-gluon
interaction instead of the full {\abbrev SM}, the number of loops
reduces by one. For example, in leading order one obtains the tree
diagram shown in Fig.\,\ref{dia::triangle}\,(b). The corresponding
expression for the partonic cross section is
\begin{equation}
  \sigma_{\mbox{\tiny LO}}(gg\to H) = 
  {\GFermi\MHiggs^2\alpha_s^2\over 288\sqrt{2}
    \pi}\,\delta(1-z)
\end{equation}
and coincides with the limit $\tau\to \infty$ of Eq.\,(\ref{eq::lo}), of
course.

Going to higher order, one has to compute virtual corrections to diagram
Fig.\,\ref{dia::triangle}\,(b). However, these will contain infra-red
and collinear divergences. The infra-red divergences will be canceled
when adding the corrections from real radiation of quarks and gluons,
but the sum will still have collinear divergences. The latter will
disappear only when the Altarelli-Parisi splitting functions are taken
into account up to the appropriate order~\cite{splitting}.  At the
next-to-leading order, a full calculation has been carried out in
\cite{dawson}. The correction terms increase the cross section by about
a factor of 1.5 to 2 in the relevant Higgs mass range.

At {\abbrev NNLO}, only a few ingredients for the full answer are
available: In \cite{schmidt} the one-loop amplitude for the radiation of
a single quark or gluon in gluon-gluon, gluon-quark, and quark-quark
fusion was obtained, and \cite{kauffman} contains the tree-level
amplitude for the double-emission of gluons and quarks in these
reactions.  Both of these results still have to be integrated over the
corresponding two- and three-particle phase space, which certainly is a
rather non-trivial task by itself.

In this paper, we want to add the virtual two-loop corrections to the
list of available knowledge at {\abbrev NNLO}.  Together with the
phase-space integrated expressions for the real radiation and the
Altarelli-Parisi splitting functions\cite{splitting}, one will then be
able to arrive at a finite result for the partonic cross section. In
order to obtain a physically accessible quantity, one needs to fold this
partonic result with the corresponding parton distribution functions.
However, they have yet to be evaluated to the appropriate order in
$\alpha_s$ (see, e.g.~\cite{moch}).


\section{Calculation of the two-loop diagrams}

\begin{figure}
  \begin{center}
    \leavevmode
    \begin{tabular}{lll}
      (a) & (b) & (c) \\
      \epsfxsize=4.cm
      \epsffile[85 235 490 490]{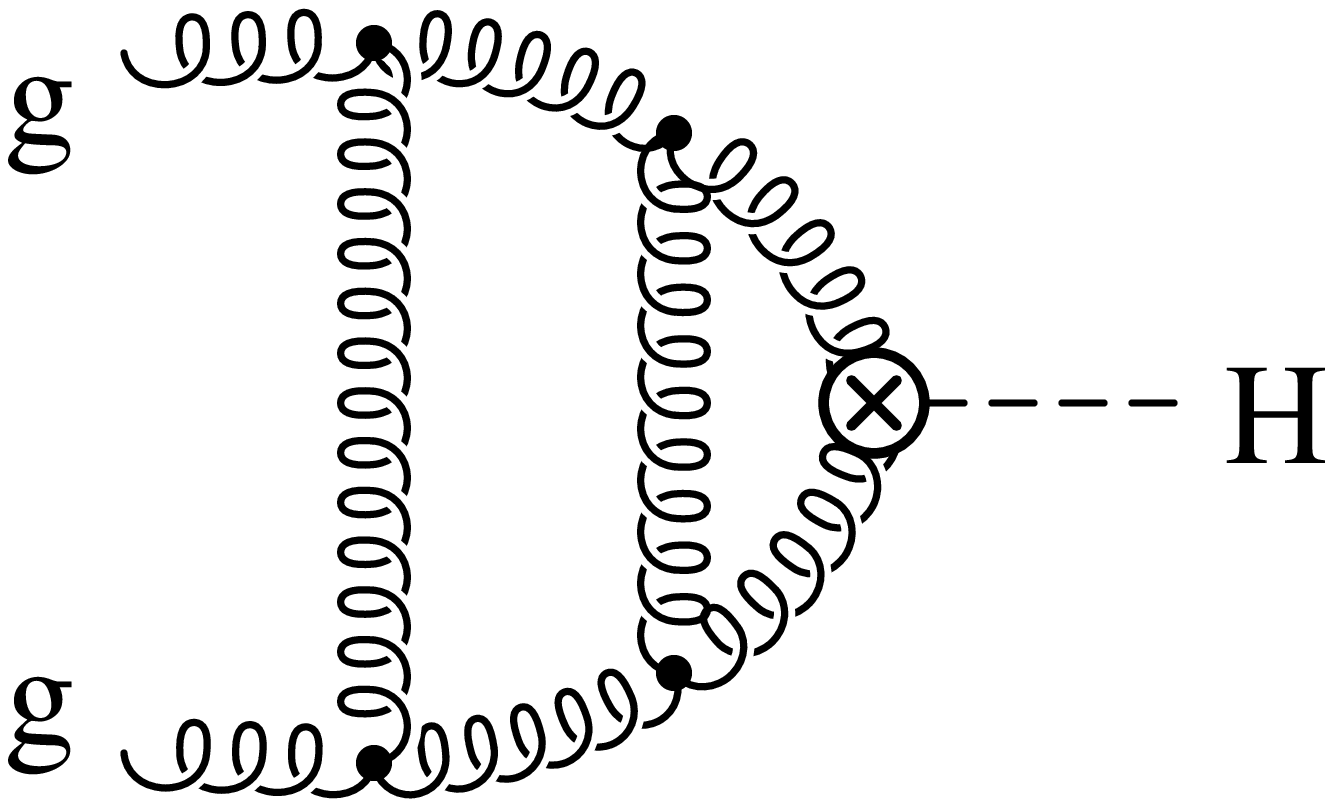} &
      \epsfxsize=4.cm
      \epsffile[85 235 490 490]{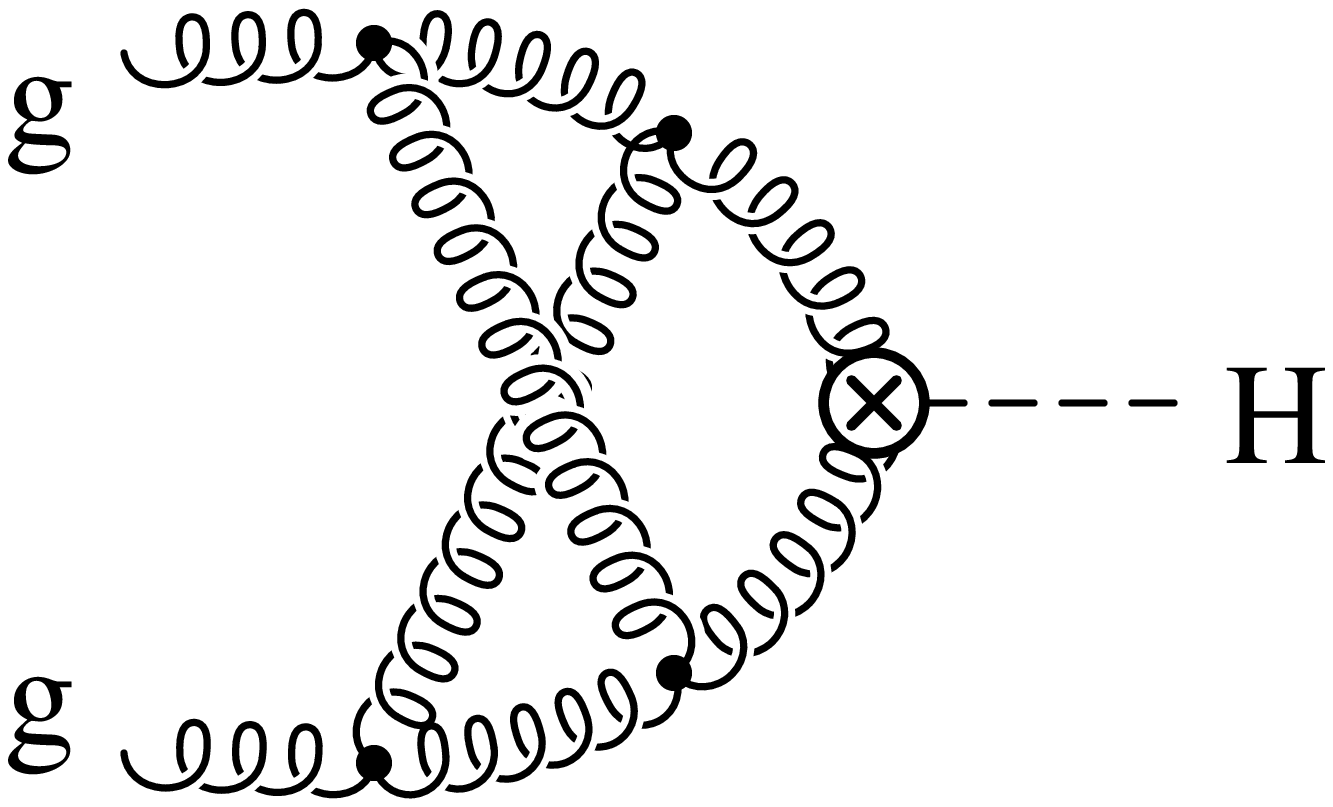} &
      \epsfxsize=4.cm
      \epsffile[85 235 490 490]{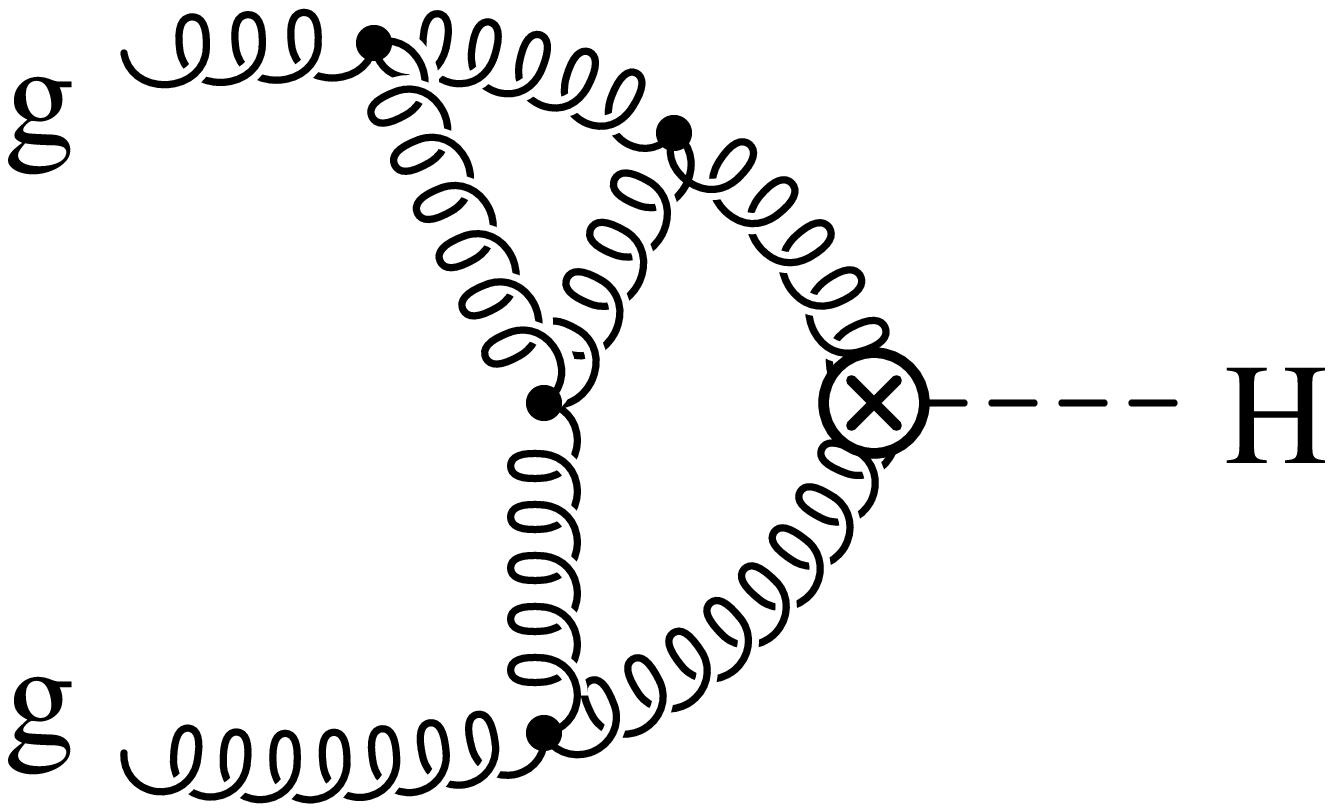}\\[1em]
        (d) & (e) & (f)\\
      \epsfxsize=4.cm
      \epsffile[85 235 490 490]{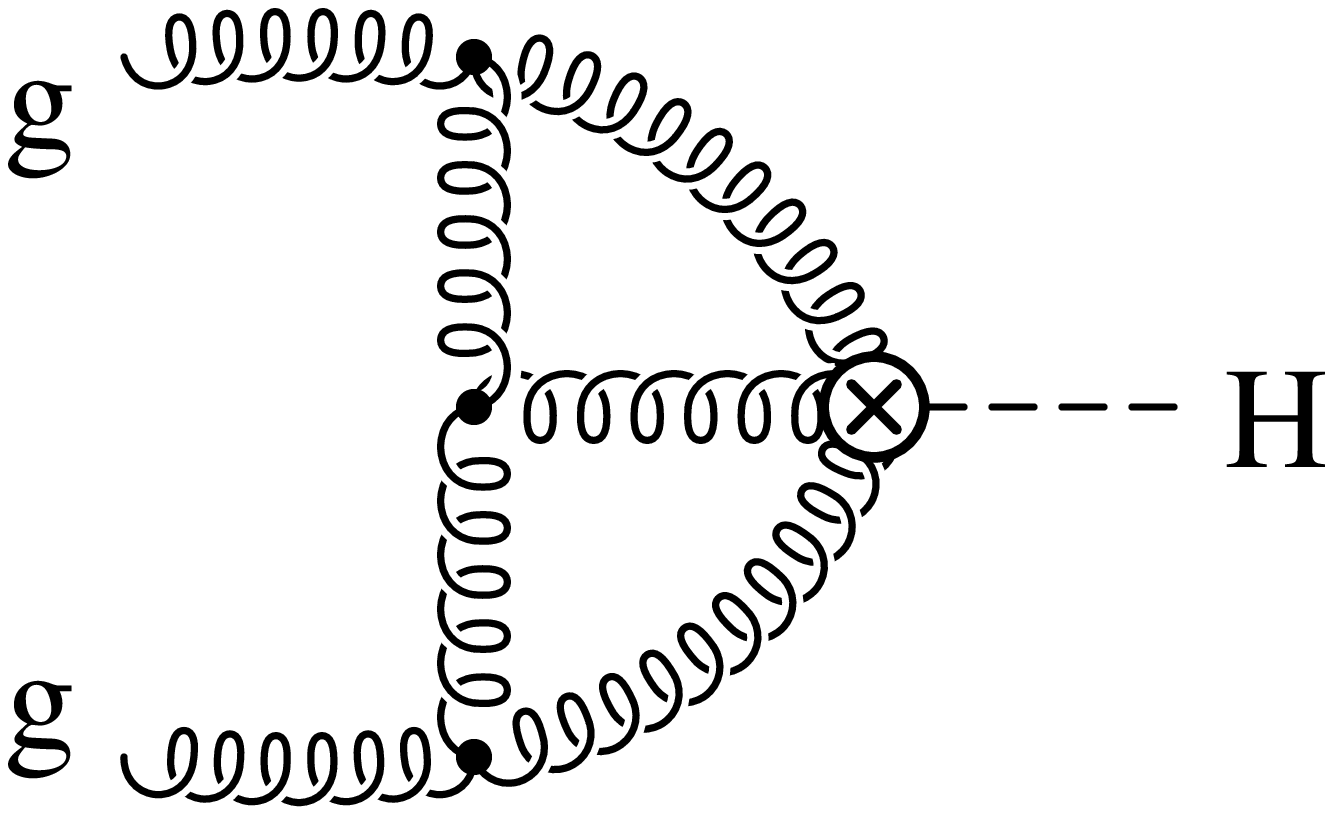} &
      \epsfxsize=4.cm
      \epsffile[85 235 490 490]{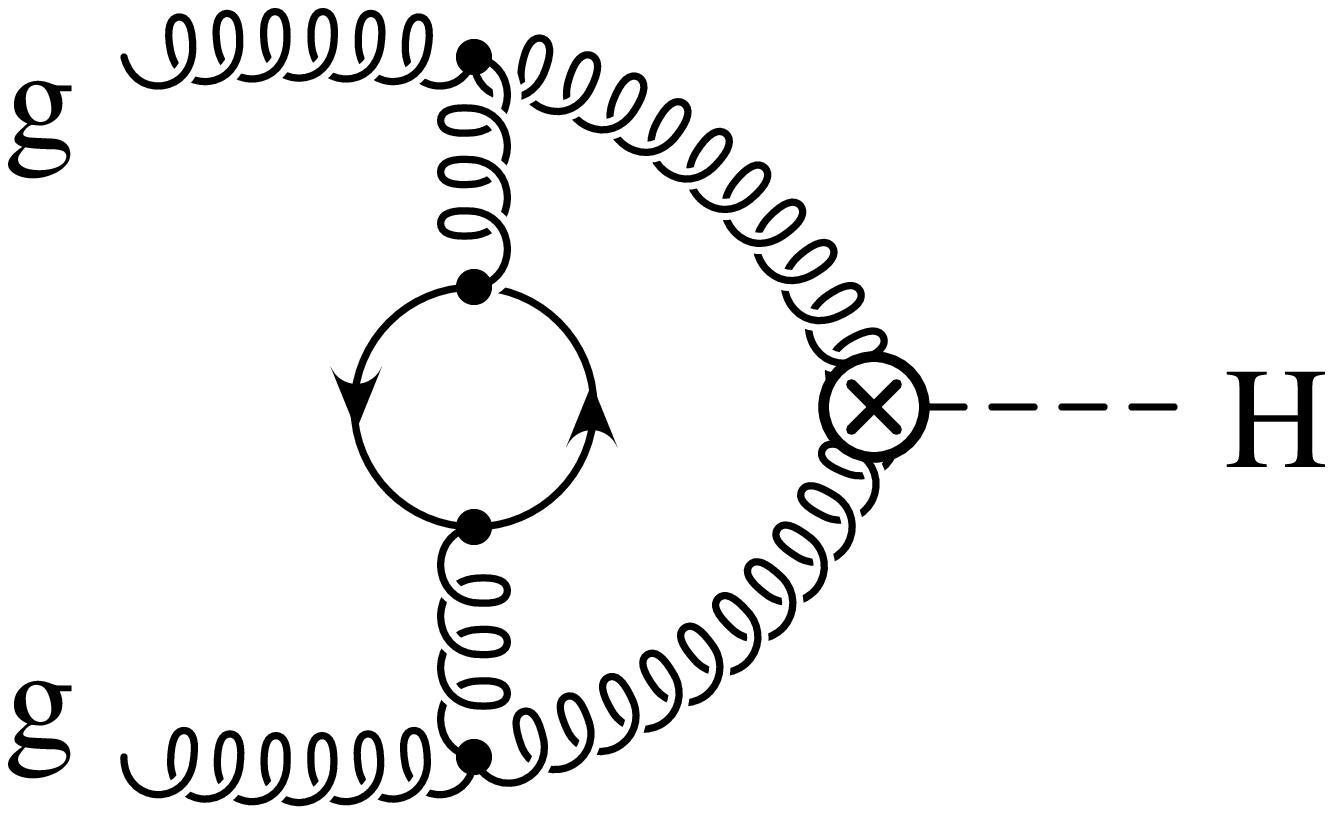} &
      \epsfxsize=4.cm
      \epsffile[85 235 490 490]{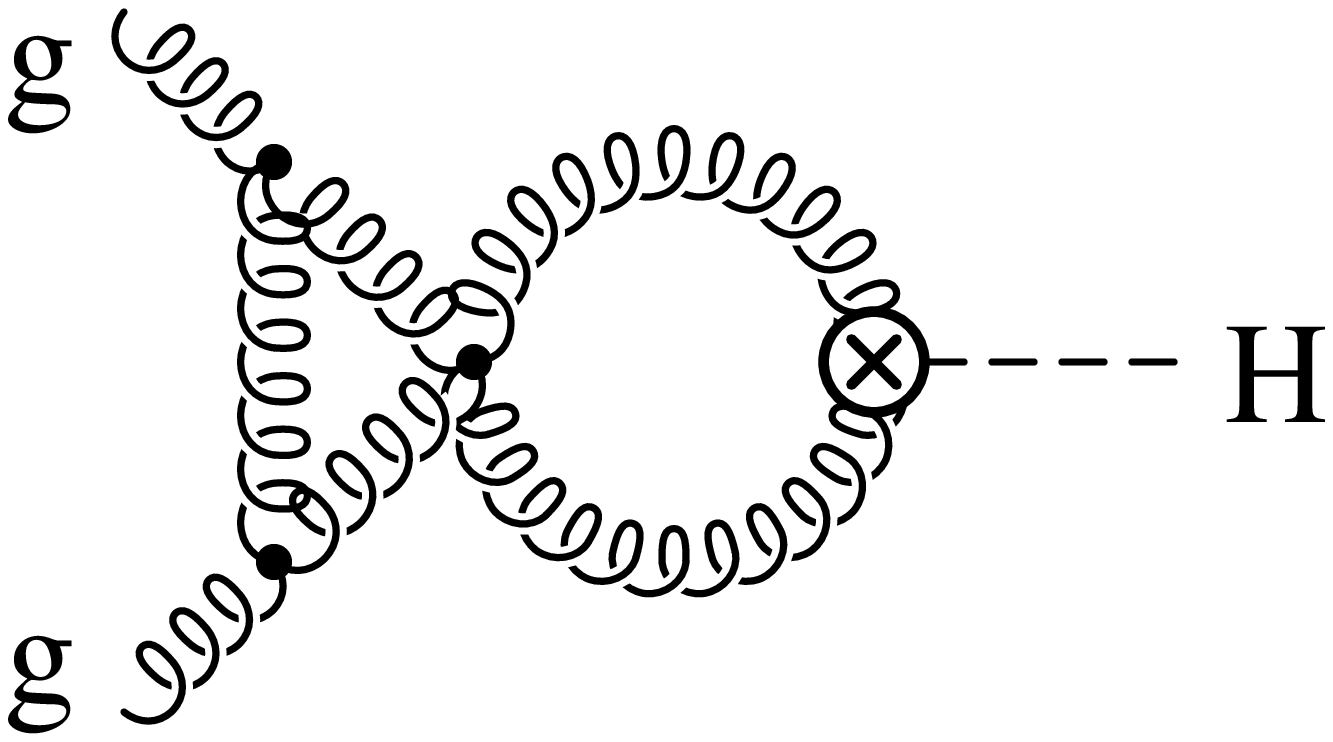}
    \end{tabular}
    \parbox{14.cm}{
    \caption[]{\label{fig::2lvirt}\sloppy
      Sample two-loop diagrams contributing to $gg\to H$ at {\abbrev
        NNLO}.  The right vertex stands for the effective coupling of
      the Higgs particle to gluons. The bubble in diagram (e) represents
      any quark except for the top quark.  }}
  \end{center}
\end{figure}

Examples of diagrams contributing to the virtual two-loop corrections to
the process $gg\to H$ are shown in Fig.\,\ref{fig::2lvirt}. The Higgs
boson couples to the effective vertex resulting from the Lagrangian of
Eq.\,(\ref{eq::leff}).  The two gluons are on-shell ($p_1^2 = p_2^2 =
0$), which is why --- after extraction of the tensor structure --- the
diagrams depend only on one kinematic variable, $(p_1 + p_2)^2 = q^2 =
M_\mathrm{H}^2$.

In addition to their topology and the power of the denominators, the
resulting integrals can be classified by the power of irreducible
numerators, i.e.\ invariants of momenta that cannot be expressed in terms
of denominators.\footnote{There is a freedom in choosing the
  specific invariants, of course.}  In
\cite{gonsalves,kramervanNeerven}, the integrals with unit (or zero)
power of denominators and low powers of irreducible numerators were
evaluated using Feynman parameterization and dispersion techniques.

In \cite{DavOs} recurrence relations based on the integration-by-parts
algorithm \cite{IP} for the planar two-loop integrals were derived,
reducing them to convolutions of one-loop integrals.  These relations
allow to compute any such planar diagram with arbitrary powers of the
denominators and irreducible numerators.

In our case, however, we also need to compute non-planar diagrams, e.g.\ 
Fig.\,\ref{fig::2lvirt}\,(b).  For this reason, we follow an algorithm
that has recently been published by Baikov and Smirnov \cite{BSvert}. It
relates the recurrence relations for $l$-loop integrals with $n+1$
external legs to the ones for ($l+1$)-loop integrals with $n$ external
legs. Here we have $n=l=2$, and thus the massless two-loop vertex
diagrams of Fig.\,\ref{fig::2lvirt} are mapped onto massless three-loop
two-point functions.  The algorithm to compute the latter ones is known
\cite{IP} and implemented in the computer program {\code MINCER}
\cite{MINCER}, written in {\code FORM} \cite{FORM}.  Following the
recipe of \cite{BSvert}, we modified the {\code MINCER} routines such
that they are applicable to the class of two-loop three-point functions
at hand.  For the generation of the diagrams we used {\code
  QGRAF}~\cite{qgraf} as integrated in the program package {\code
  GEFICOM}~\cite{geficom}\footnote{I acknowledge the kind permission by
  the authors of {\scriptsize\sf GEFICOM} to use this program.}.

The only integral that can not be reduced to convolutions of one-loop
integrals in this approach is the non-planar one with all propagators
appearing in single power, and with numerator equal to one. However, the
result for this integral is known as an expansion in $\epsilon$ up to
its finite part \cite{gonsalves}.

As a check of our setup we re-did the calculation of the
electro-magnetic quark form factor in {\abbrev QCD} to two loops and
found full agreement with \cite{kramervanNeerven}. We performed this
calculation in a general $R_\xi$ gauge and explicitly checked its gauge
parameter independence in this way.  We also computed the two-loop
three-gluon vertex in $R_\xi$ gauge with two gluons on-shell and found
agreement with \cite{DavOs,DavOsTar,nowak}.  Finally, the calculation of
the present paper was also performed in $R_\xi$ gauge and we verified
that the gauge parameter dependence disappears in the sum of all
diagrams.


\section{Results}

The virtual cross section for the process $gg\to H$ can be written as
\begin{equation}
\sigma_\mathrm{virt} = 4\pi{\MHiggs^2\over v^2}
  \,\delta\!\left(1-z\right)\,\left({C_1\over
  1-\beta(\norm\alpha_s)/\ep}\right)^2
  {1\over 256(1-\ep)^2} \sum_{\rm pol}\left|{\cal M}\right|^2\,,
\end{equation}
where
\begin{equation}
  {\cal M} = 
  \varepsilon^{a,\mu}(p_1,\lambda_1) \varepsilon^{b,\nu}(p_2,\lambda_2)
  \cala_{\mu\nu}^{ab}(p_1,p_2)\,,
\end{equation}
and $C_1$ is the coefficient function given in Eq.\,(\ref{eq::coefc}).
The factor $1/256/(1-\ep)^2$ comes from the average over initial state
polarizations and color.
The kinematical constraint on the momenta is
\begin{equation}
p_1 + p_2 = q\,,\qquad q^2 = \MHiggs^2\,,\qquad p_1^2 = p_2^2 = 0\,.
\end{equation}
$\varepsilon^{a,\mu}(p,\lambda)$ is the polarization vector of a gluon
with momentum $p$ and polarization $\lambda$, and $a,b$, respectively
$\mu,\nu$ are the {\abbrev SU(3)}-color and the Lorentz indices.  Adopting a
covariant $R_\xi$ gauge, the polarization vectors obey the relation
\begin{equation}
\sum_\lambda \varepsilon^{a\ast}_\mu(p,\lambda) 
\varepsilon^{b}_\nu(p,\lambda) =
-g_{\mu\nu}\delta^{ab}\,.
\end{equation}

The tensor $\cala_{\mu\nu}^{ab}$ may be written as
\begin{equation}
\cala_{\mu\nu}^{ab} = {\delta^{ab}\over 2\MHiggs^2}\left[
  a\,(\ponetwo)g_{\mu\nu} + b\,p_{1\nu}p_{2\mu} +
 c\,p_{1\mu}p_{2\nu} + d\,p_{1\mu}p_{1\nu} + e\,p_{2\mu}p_{2\nu}\right]\,.
\label{eq::tensor}
\end{equation}
The last two terms seem to violate gauge invariance. However, to lowest
order they vanish, and starting from next-to-leading order, their
contribution to the squared matrix element gets canceled by the
diagrams with ghosts in the initial state.  We explicitly checked this
cancelation by computing these ghost diagrams to the corresponding
order.  Further, we have
\begin{equation}
a = -b\,,
\end{equation}
and thus the $c$ term in (\ref{eq::tensor}) does not contribute to the
total rate.  Therefore we will only quote the result for $a$ in the
following.  Its loop expansion can be written as
\begin{equation}
  a = 1
  + \apib\, a^{(1)}
  + \left(\apib\right)^2 a^{(2)} + \ldots\,.
  \label{eq::a}
\end{equation}
The first order correction has been computed in \cite{dawson} up to the
finite part. Since it involves a second order pole in
$\epsilon$ and we are also interested in the finite part of its square,
we need its expansion up to $\ep^2$:
\begin{equation}
  \begin{split}
    a^{(1)} = \norm\left(-{\mu^2\over \MHiggs^2}\right)^\ep
    \bigg\{
    -{3\over 2\ep^2}
    + {3\over 4}\,\zeta_2 
    + \ep\,\bigg(
    - {3\over 2} 
    + {7\over 2}\,\zeta_3
    \bigg)
    + \ep^2\,\bigg(
    - {9\over 2}
    + {141\over 32}\,\zeta_4
    \bigg)
    \bigg\} + \order{\ep^3}\,,
  \end{split}
  \label{eq::a1}
\end{equation}
where $\mu$ is the 't\,Hooft mass, and $\zeta_n \equiv \zeta(n)$ is
Riemann's zeta function ($\zeta_2 = \pi^2/6$; $\zeta_3
\approx 1.20206$; $\zeta_4 = \pi^4/90$). Furthermore, it is understood
that $(-1)^\ep \equiv \exp(+i\ep\pi)$.

To this we add the second order correction:
\begin{equation}
  \begin{split}
    a^{(2)} &= \norm^2\left(-{\mu^2\over
        \MHiggs^2}\right)^{2\ep}
    \bigg\{{9\over 8\ep^4} + {1\over
      \ep^3}\,\bigg[ -{33\over 32} + {1\over 16}\,n_l \bigg] + {1\over
      \ep^2}\,\bigg[ -{67\over 32} - {9\over 16}\,\zeta_2 + {5\over
      48}\,n_l \bigg] \\& + {1\over \ep}\,\bigg[ {17\over 12} +
    {99\over 32}\,\zeta_2 - {75\over 16}\,\zeta_3 + n_l\,\bigg(
    -{19\over 72} - {3\over 16}\,\zeta_2 \bigg) \bigg] \\& + {5861\over
      288} + {201\over 32}\,\zeta_2 + {11\over 16}\,\zeta_3 -
    {189\over 32}\,\zeta_4 + n_l\,\bigg[ -{605\over 216} - {5\over
      16}\,\zeta_2 - {7\over 8}\,\zeta_3 \bigg] \bigg\}\,,
  \end{split}
  \label{eq::a2}
\end{equation}
where, as before, $n_l$ is the number of light quark flavors, $n_l = 5$.

Ultra-violet renormalization of the strong coupling constant is given by
$\alpha_s^\bare = Z_\alpha(\norm\alpha_s)\cdot\alpha_s$, where $Z_\alpha$
is related to the $\beta$ function of Eq.~(\ref{eq::beta}) through
\begin{equation}
\alpha_s\doverd{\alpha_s}\ln Z_\alpha(\alpha_s) = {\beta(\alpha_s)\over \ep -
  \beta(\alpha_s)}\,.
\end{equation}


\section{Ratio of time-like to space-like form factor}
Both as a check and as an estimate on the magnitude of the corrections,
we may consider the ratio of the time-like to the space-like form
factor. It is free from infra-red singularities and contains the
presumably most significant contributions stemming from the analytic
continuation of the factor $(\mu^2/(-q^2))^\ep$ from space-like to
time-like values of $q^2$.

As was shown in~\cite{sterman} for the quark form factor in {\abbrev
  QCD}~\cite{kramervanNeerven}, this ratio can be derived up to
$\order{\alpha_s^2}$ from the {\it one}-loop terms of the form factor by
combining them with a known two-loop anomalous dimension.  In the case
of $gg\to H$ this anomalous dimension is given by $9/4$ times the one
given in~\cite{sterman}.  Following the derivation of~\cite{sterman} and
setting $\mu^2 = \MHiggs^2$, we arrive at the following expression:
\begin{equation}
\begin{split}
\left|{a(\MHiggs^2)\over a(-\MHiggs^2)}\right|^2 &=
  1+ {3\over 2}\,\pi^2\,{\alpha_s(\MHiggs^2)\over \pi} +
  \left({\alpha_s(\MHiggs^2)\over \pi}\right)^2
  \left( {3\over 4}\,\pi^4 + {67\over 8}\,\pi^2 - {5\over
  12}\,\pi^2\,n_l\right) + \order{\alpha_s^3}\\
&\approx 1 + 14.8 {\alpha_s(\MHiggs^2)\over \pi} + 153.2
  \left({\alpha_s(\MHiggs^2)\over \pi}\right)^2 + \order{\alpha_s^3}\\
&\approx 1 + 0.528 + 0.172\approx 1.700\,,
\label{eq::ratiogam}
\end{split}
\end{equation}
where we inserted $n_l = 5$ in the second line. The third line displays
separately the {\abbrev LO, NLO} and the {\abbrev NNLO} contribution, as
well as their sum, for $\alpha_s(\MHiggs^2) = 0.112$.
\eqn{eq::ratiogam} fully agrees with a direct evaluation of the ratio
using the expressions (\ref{eq::a}), (\ref{eq::a1}), and (\ref{eq::a2})
for $a$. This provides a check on the terms $\propto
\alpha_s^{n}/\ep^{2n-k}$ ($k=0,1,2$) of our result.  The numerical value
of the {\abbrev NLO} correction in \eqn{eq::ratiogam} reflects the
largeness of the full {\abbrev NLO} terms as obtained in \cite{dawson}.
The number for the {\abbrev NNLO} corrections gives some hope towards a
certain degree of convergence for the perturbative series of the full
result for $\sigma(gg\to H)$.  Concluding this section, let us note that
an interesting extension of this discussion could be the resummation of
the leading terms along the lines of \cite{magnea}.


\section{Conclusions and Outlook}
We used the recently introduced method of \cite{BSvert} in order to
calculate the {\abbrev NNLO} virtual corrections to the production cross
section of Higgs bosons in gluon fusion.  The result is a gauge
invariant component of the full cross section.  The next step towards a
complete answer for the {\abbrev NNLO} rate is to integrate the
squared amplitudes for the real radiation processes over the phase
space. This is work in progress~\cite{kilgore}.
Finally, one has to convolute the full partonic cross section with the
parton distribution functions.
Their evaluation to the relevant order is therefore certainly a very
important task.


\section*{Acknowledgments}
I would like to thank the High Energy/Nuclear Theory group at BNL, in
particular A.~Czarnecki, S.~Dawson, W.~Kilgore, and W.~Vogelsang, for
encouragement and valuable discussions.  Furthermore, I acknowledge
discussions and comments by K.~Melnikov and T.~Seidensticker.  This work
was supported by the {\it Deutsche Forschungsgemeinschaft}.

  
\def\app#1#2#3{{\it Act.~Phys.~Pol.~}{\bf B #1} (#2) #3}
\def\apa#1#2#3{{\it Act.~Phys.~Austr.~}{\bf#1} (#2) #3}
\def\annphys#1#2#3{{\it Ann.~Phys.~}{\bf #1} (#2) #3}
\def\cmp#1#2#3{{\it Comm.~Math.~Phys.~}{\bf #1} (#2) #3}
\def\cpc#1#2#3{{\it Comp.~Phys.~Commun.~}{\bf #1} (#2) #3}
\def\epjc#1#2#3{{\it Eur.\ Phys.\ J.\ }{\bf C #1} (#2) #3}
\def\fortp#1#2#3{{\it Fortschr.~Phys.~}{\bf#1} (#2) #3}
\def\ijmpc#1#2#3{{\it Int.~J.~Mod.~Phys.~}{\bf C #1} (#2) #3}
\def\ijmpa#1#2#3{{\it Int.~J.~Mod.~Phys.~}{\bf A #1} (#2) #3}
\def\jcp#1#2#3{{\it J.~Comp.~Phys.~}{\bf #1} (#2) #3}
\def\jetp#1#2#3{{\it JETP~Lett.~}{\bf #1} (#2) #3}
\def\mpl#1#2#3{{\it Mod.~Phys.~Lett.~}{\bf A #1} (#2) #3}
\def\nima#1#2#3{{\it Nucl.~Inst.~Meth.~}{\bf A #1} (#2) #3}
\def\npb#1#2#3{{\it Nucl.~Phys.~}{\bf B #1} (#2) #3}
\def\nca#1#2#3{{\it Nuovo~Cim.~}{\bf #1A} (#2) #3}
\def\plb#1#2#3{{\it Phys.~Lett.~}{\bf B #1} (#2) #3}
\def\prc#1#2#3{{\it Phys.~Reports }{\bf #1} (#2) #3}
\def\prd#1#2#3{{\it Phys.~Rev.~}{\bf D #1} (#2) #3}
\def\pR#1#2#3{{\it Phys.~Rev.~}{\bf #1} (#2) #3}
\def\prl#1#2#3{{\it Phys.~Rev.~Lett.~}{\bf #1} (#2) #3}
\def\pr#1#2#3{{\it Phys.~Reports }{\bf #1} (#2) #3}
\def\ptp#1#2#3{{\it Prog.~Theor.~Phys.~}{\bf #1} (#2) #3}
\def\ppnp#1#2#3{{\it Prog.~Part.~Nucl.~Phys.~}{\bf #1} (#2) #3}
\def\sovnp#1#2#3{{\it Sov.~J.~Nucl.~Phys.~}{\bf #1} (#2) #3}
\def\tmf#1#2#3{{\it Teor.~Mat.~Fiz.~}{\bf #1} (#2) #3}
\def\tmp#1#2#3{{\it Theor.~Math.~Phys.~}{\bf #1} (#2) #3}
\def\yadfiz#1#2#3{{\it Yad.~Fiz.~}{\bf #1} (#2) #3}
\def\zpc#1#2#3{{\it Z.~Phys.~}{\bf C #1} (#2) #3}
\def\ibid#1#2#3{{ibid.~}{\bf #1} (#2) #3}


\begin{thebibliography}{99}
\bibitem{BSvert} P.A.~Baikov and V.A.~Smirnov, \plb{477}{2000}{367}.
\bibitem{lepewwg} See \verb$http://lepewwg.web.cern.ch/LEPEWWG/$ for updates.
\bibitem{spira}
  M.~Spira, A.~Djouadi, D.~Graudenz, and P.M.~Zerwas,
  \npb{453}{1995}{17};  M.~Spira, \fortp{46}{1998}{203}.
\bibitem{lo} F.~Wilczek, \prl{39}{1977}{1304};
  J.~Ellis, M.~Gaillard, D.~Nanopoulos, and C.~Sachrajda,
    \plb{83}{1979}{339};
    H.~Georgi, S.~Glashow, M.~Machacek, and D.~Nanopoulos,
    \prl{40}{1978}{692}; T.~Rizzo, \prd{22}{1980}{178}.
\bibitem{dawson} A.~Djouadi, M.~Spira, and P.M.~Zerwas, \plb{264}{1991}{440};\\
  S.~Dawson, \npb{359}{1991}{283}.

\bibitem{kauffman1} S.~Dawson and R.P.~Kauffman, \prd{49}{1994}{2298}.
\bibitem{CheKniSte97} K.G.~Chetyrkin,
  B.A.~Kniehl, and M.~Steinhauser, \npb{510}{1998}{61}.
\bibitem{c1a3} K.G.~Chetyrkin, B.A.~Kniehl, and M.~Steinhauser,
  \prl{79}{1997}{353};
  M.~Kr\"amer, E.~Laenen, and M.~Spira, \npb{511}{1998}{523}.
\bibitem{spiridonov}
  V.P.~Spiridonov, Rep.~No.~{\abbrev INR P-0378} (Moscow, 1984);
  V.P.~Spiridonov and K.G.~Chetyrkin , \yadfiz{47}{1988}{818}.
\bibitem{beta4} T.~van~Ritbergen, J.A.M.~Vermaseren, and S.A.~Larin,
    \plb{400}{1997}{379}.
\bibitem{splitting}
  G.~Curci, W.~Furmanski, and R.~Petronzio, \npb{175}{1980}{27}.
\bibitem{schmidt} C.R.~Schmidt, \plb{413}{1997}{391}.
\bibitem{kauffman} S.~Dawson and R.P.~Kauffman, \prl{68}{1992}{2273}; 
  R.P.~Kauffman, S.V.~Desai, and D.~Risal,
  \prd{55}{1997}{4005}; (E) \ibid{D 58}{1998}{119901}.
\bibitem{moch} S.~Moch and J.A.M.~Vermaseren, proc.\ {\it
    Zeuthen Workshop on Elementary Particle Theory: Loops and Legs in
    Quantum Field Theory}, K\"onigstein-Wei\ss{}ig, Germany, 9-14 Apr
    2000; hep-ph/0006053.
\bibitem{gonsalves} R.J.~Gonsalves, \prd{28}{1983}{1542}.
\bibitem{kramervanNeerven} G. Kramer and B. Lampe,
  \zpc{34}{1987}{497}; (E)~\ibid{42}{1989}{504};\\
  T. Matsuura, S.C. van der Marck and W.L. van Neerven,
  \npb{319}{1989}{570}.  
\bibitem{DavOs} A.I.~Davydychev and P.~Osland, \prd{59}{1998}{014006}.
\bibitem{IP} F.V.~Tkachov, \plb{100}{1981}{65};\\
     K.G.~Chetyrkin and F.V.~Tkachov, \npb{192}{1981}{159}.
\bibitem{MINCER} S.A.~Larin, F.V.~Tkachov, and J.A.M.~Vermaseren,
  Rep.~No.~{\abbrev NIKHEF-H/91-18} (Amsterdam, 1991).
\bibitem{FORM} J.A.M.~Vermaseren, Symbolic Manipulation with {\code
FORM}, {\abbrev CAN} (1991).
\bibitem{qgraf} P.~Nogueira, \jcp{105}{1993}{279}.
\bibitem{geficom} K.G.~Chetyrkin and M.~Steinhauser, unpublished.
\bibitem{DavOsTar} A.I.~Davydychev, P.~Osland, and O.V.~Tarasov,
  \prd{54}{1996}{4087}; (E)~\ibid{59}{1999}{109901}.  
\bibitem{nowak}  M.A.~Nowak, M.~Prasza\l{}owicz, and W.~S\l{}omi\'nski,
  \annphys{166}{1986}{443}.  
\bibitem{sterman} L.~Magnea and G.~Sterman, \prd{42}{1990}{4222}.
\bibitem{magnea} L.~Magnea, Rep.~No.~{\abbrev DFTT 22/00} 
  (Torino, 2000); hep-ph/0006255.
\bibitem{kilgore} R.V.~Harlander and W.B.~Kilgore, in preparation.

\end{thebibliography}
\end{document}